\documentclass[prc,twocolumn,english,showpacs]{revtex4}
\usepackage[T1]{fontenc}
\usepackage[latin1]{inputenc}
\usepackage{babel}
\usepackage{graphics}
\usepackage{comment}

\makeatletter

\renewcommand {\vec}[1]{\mbox{\boldmath $#1$}}

\makeatother

\begin{document}

\title{Three-body continuum discretization in a basis of transformed harmonic
oscillator states}

\author{M. Rodr\'{\i}guez-Gallardo$^{1,2}$, J.M.~Arias$^1$,
J.~G\'omez-Camacho$^1$, A.M. Moro$^1$, I.J. Thompson$^2$, and J.A.
Tostevin$^2$}

\affiliation {$^1$ Departamento de F\'{\i}sica At\'omica,
Molecular y Nuclear, Facultad de F\'{\i}sica, \\ Universidad de
Sevilla, Apartado 1065, 41080 Sevilla, Spain} \affiliation{$^2$
Department of Physics, University of Surrey, Guildford GU2 7XH,
United Kingdom.}
\date{\today}

\begin{abstract}
The inclusion of the continuum in the study of weakly-bound
three-body systems is discussed. A transformed harmonic oscillator
basis is introduced to provide an appropriate discrete and finite
basis for treating the continuum part of the spectrum. As examples
of the application of the method the strength functions
corresponding to several operators that couple the ground state to
the continuum are investigated, for $^6$He, and compared with
previous calculations. It is found that the energy moments of
these distributions are accurately reproduced with a small basis
set.
\end{abstract}
\pacs{21.45.+v,21.10.-k, 27.20.+n, 03.65.Ca}
\maketitle

\section{Introduction}

The general solution of a quantum mechanical Hamiltonian problem
containing a time-independent potential gives rise to both bound
and unbound eigenstates. Usually the Hamiltonian of the system has
a finite number of bound eigenstates while the unbound ones form a
continuum. For the description of nuclei in the stability valley
usually only the bound eigenstates are considered. However, the
development of radioactive nuclear beam facilities has allowed the
study of nuclei far from the line of stability, bringing to the
fore new nuclear structure problems. One of the main topics in
recent years has been the study of halo nuclei \cite{Han95,Jen04}.
These are weakly-bound, spatially extended systems, typically
comprising a core and one or two valence nucleons.

A particularly interesting example of exotic systems is that of
Borromean nuclei, i.e., three-body composite systems with no
binary bound states. These nuclei have deserved special attention
because their loosely-bound nature reflects a delicate interplay
between two- and three-body forces, thus constituting a challenge
to existing theories, and a motivation for the development of new
ones. Even today, the detailed structure of the continuum spectrum
of these systems is not fully understood, partially due to the
ambiguities associated with the underlying forces between the
constituents. Due to their low binding energy, halo nuclei are
easily broken up in the nuclear and Coulomb field of the target.
Therefore few-body reaction theories, developed to extract
reliable information from experimental data of reactions involving
loosely bound systems, have to include, as an essential
ingredient, a realistic description of the continuum part of the
spectrum.

From the theoretical point of view, the treatment of  reactions
involving loosely bound systems deals with the complication that
breakup states are not square-normalizable. A convenient method to
circumvent this problem is to replace the states in the continuum
by a finite set of normalized states, thus providing a discrete
basis that, hopefully, can be truncated to a small number of
states and give a good description of the continuum. Several
prescriptions to construct a discrete basis have been proposed.
For two-body composite systems, where true continuum states are
easily calculated, one can use a discretization procedure in which
the continuum spectrum is truncated at a maximum excitation energy
and divided into energy intervals. For each interval, or bin, a
normalizable state is constructed by superposition of scattering
states within that bin interval.  The method, known as the
Continuum Discretized Coupled Channels (CDCC) method
\cite{Yah86,Aus87}, has been very useful in the description of
elastic and breakup observables in reactions involving weakly
bound two-body projectiles. An alternative method to obtain a
discrete representation of the continuum spectrum is to
diagonalize the two-body Hamiltonian in a complete set of $L^2$
functions such as Gaussians \cite{Mat03,Mac87a} or Laguerre
functions \cite{Bra95,Kur82,Kur91}. This method has the appealing
feature of being readily applicable to three-body systems, in
which case the Hamiltonian is diagonalized in a complete set of
square-integrable functions for the 3-body Hilbert space. Several
applications of this method can be found in the literature, for
both structure \cite{Hiy03} and reaction problems \cite{Mat04}. In
the latter case, the method constitutes a natural extension of the
CDCC formalism for reactions with three-body projectiles.

When the ground state wave function is already known, a useful
procedure to obtain a discrete representation for scattering
states consists of performing a Local Scale Transformation (LST)
\cite{Pet81} that transforms the ground state wave function of the
system into the ground state of a Harmonic Oscillator (HO)
\cite{Per01a,Per01b}. Once the LST is obtained, the HO basis can
be transformed by the inverse LST to a discrete basis in the
physical space. The functions in the Transformed Harmonic
Oscillator (THO) basis are not eigenfunctions of the Hamiltonian
(except for the ground state) but the Hamiltonian can be
diagonalized in an appropriate truncated basis to produce
approximate eigenvalues and eigenfunctions. This method has been
shown to be useful for describing the two-body continuum in both
structure \cite{Per01a,Per01b,Rod04} and scattering
\cite{Mor02,Mar02} problems. In particular, it was shown that
global structure functions, related to the coupling to the
continuum, such as strength functions, are very accurately
described using a relatively small THO basis.

In this work we extend the THO formalism presented in
\cite{Per01a,Per01b} to treat the three-body continuum. In
particular we apply the method to the Borromean nucleus $^6$He.
This is a very weakly bound system with a well developed
structure, of an  $\alpha$ cluster and two valence neutrons. Most
of our knowledge of this nucleus comes from the analysis of
reactions where secondary beams of $^6$He collide with stable
nuclei. These experiments have been performed with both light
\cite{Aks03,Ege02} and heavy targets
\cite{Aum98,Aum99,Agu01,Agu00,Kak03}, and at low and high
energies, providing a rich variety of data which can be used to
benchmark reaction and structure models.

Theoretically, $^6$He has been the object of many studies, using
three-body methods \cite{Zhu93,Hi95,Dan97,Myo01,Fed03,Gar04},
cluster-orbital shell models \cite{Su91,Fu94}, no-core microscopic
shell models \cite{Na96} and microscopic cluster models, and for
various effective nucleon-nucleon interactions \cite{Cso93,Wur97}.
In the present work, the three-body equations are solved according
to the formalism described in Ref.~\cite{Zhu93}. In this
reference, several three-body methods are compared and applied to
the ground state properties of the Borromean nuclei $^{6}$He and
$^{11}$Li. One of these approaches, discussed in \cite{Zhu93},
consists of a solution of the Faddeev equations in configuration
space. In the Faddeev formalism, the total wave function for a
three-body system is expressed as a superposition of three terms,
one for each Jacobi configuration. This wave function, which
contains both scattering and rearrangement channels, is obtained
by solving a set of coupled integro-differential equations. A
practical way to solve the problem is to express the three-body
wave function in terms of hyperspherical coordinates. For
convenience, scattering states are expanded in terms of a discrete
basis. In Ref. \cite{Zhu93} a family of Gauss-Laguerre functions
was used for this purpose. In the present work, we choose a THO
basis obtained from a 3-body calculation of the $^{6}$He nucleus.

The method we present can also be used as an input for reaction
theories. However, in the present paper we present only results
related to the structure of $^6$He. In particular, we investigate
strength functions for several operators that couple the $^6$He
ground state to the continuum. Results related to scattering
processes induced by $^6$He will be presented in a forthcoming
paper.

This paper is structured as follows. In Section II the THO
formalism for a three-body system is developed. In Section III the
formalism is then applied to the case of the structure of $^6$He.
Finally, Section IV summarizes and draws conclusions from this
work.

\section{Formalism of the transformed harmonic oscillator (THO) states
for a three-body continuum system}

In this Section we develop the formalism of transformed harmonic
oscillator states to be applied to three-body continuum systems.
To a large extent, this formalism is a generalization of the
method presented in \cite{Per01a,Per01b}. For such a system we
will use the hyperspherical coordinates that are obtained from the
Jacobi coordinates (see, for instance, \cite{face}). Consider
three particles of masses $\{m_1,m_2,m_3\}$ and whose locations
are defined by $\{\vec r_1,\vec r_2,\vec r_3\}$. Three sets of
Jacobi coordinates $\{\vec{x}_i,\vec{y}_i\}$, with $i=1,2,3$, can
be defined where $\{\vec{x}_i\}$ is the relative coordinate for
two particles and $\{\vec{y}_i\}$ is the coordinate of the third
particle relative to the center of mass of the two particles used
to define $\{\vec{x}_i\}$:
\begin{eqnarray}
\vec{x}_i&=&(\vec{r}_j-\vec{r}_k)\sqrt{\frac{\mu_{x_i}}{m}}, \nonumber
\\ \vec{y}_i&=&\left(\vec{r}_i-\frac{m_j\vec{r}_j+m_k\vec{r}_k}{m_j+
m_k}\right)\sqrt{\frac{\mu_{y_i}}{m}}.
\end{eqnarray}
Here $\mu_{x_i}$ is the reduced mass of the $\{j,k\}$ system and
$\mu_{y_i}$ is the reduced mass of the $(j+k)$ and particle $i$
system,
\begin{eqnarray}
\mu_{x_i}&=&\frac{m_jm_k}{m_j+m_k}, \nonumber \\
\mu_{y_i}&=&\frac{m_i(m_j+m_k)}{m_i+m_j+m_k},
\end{eqnarray}
and $m$ is an arbitrary normalization mass that we will take as
the nucleon mass. The indices $i,j,k$ run through (1,2,3) in
cyclic order.
\begin{figure}
\resizebox*{0.25\textwidth}{!}{\includegraphics{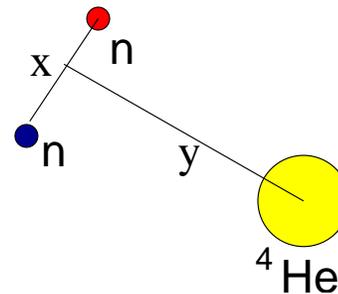}}
\vspace{-0.4cm} \caption{\label{jacobi-6he} The Jacobi
T-coordinate system used to describe the $^6$He system.}
\end{figure}

From these Jacobi coordinates, one can introduce the
hyperspherical coordinates defined by the hyperradius $\rho$ and
the hyperangle $\alpha_i$,
\begin{eqnarray}
\rho^2 &= & x_1^2+y_1^2=x_2^2+y_2^2=x_3^2+y_3^2~ , \nonumber \\
\tan{\alpha_i}&=&{x_i}/{y_i}, \label{rho}
\end{eqnarray}
with $x_i=\rho \sin \alpha_i$ and $y_i=\rho \cos \alpha_i$. Note
that the hyperangle depends on the Jacobi system selected but not
on the hyperradius.

For the three-body system of a core, considered inert and
spin-less, plus two valence particles, the total wave function is
assumed to be a product of the core intrinsic wave function
$\phi_c(\xi_c)$ and the two-valence-particle plus core relative
motion wave function $\Psi_{JM}(1,2)$. $\Psi_{JM}(1,2)$ depends on
the relative coordinates and the spins of the two particles and is
the solution of the Schr\"odinger equation
\begin{equation}
(\widehat{T}+\widehat{V}-E)\Psi(1,2)=0
\label{hamil}
\end{equation}
with
\begin{equation}
\widehat{V}=\widehat{V}_{c1}+\widehat{V}_{c2}+\widehat{V}_{12}.
\end{equation}
$\widehat{V}_{ci}$ is the interaction of the core with particle
$i$ and $\widehat{V}_{12}$ the interaction between particles $1$
and $2$.

From this point, we work within the Jacobi T-coordinate system,
shown in Fig. \ref{jacobi-6he}, and hence all subsequent formulae
refer to this coordinate set. Expanding $\Psi_{JM}(1,2)$ using
the hyperspherical harmonics (HH) referred to these coordinates
\begin{equation}
\Psi_{JM}(1,2)= \sum_{\beta} R_{\beta}(\rho)\left[
\Upsilon_{Kl}^{l_xl_y}(\Omega)\otimes X_{S}\right]_{JM},
\end{equation}
where $\rho$ is the hyperradius, $\beta\equiv\{K,l_x,l_y,l,S\}$
labels each channel, $R_{\beta}(\rho)$ is the hyperradial wave
function, $\Upsilon_{Kl}^{l_xl_y}(\Omega)$ is a hyperspherical
harmonic in the angles $\Omega=(\alpha,\hat x,\hat y)$, and
$X_{S}(1,2)$ is the total spin wave function of particles $1$ and
$2$. $\vec{J}=\vec{l}+\vec{S}$ is the total angular momentum of
the system, since the core particle is assumed spinless. The
hyperspherical harmonics are
\begin{eqnarray}
\Upsilon_{Klm_l}^{l_xl_y}(\Omega)&=&\Psi_K^{l_xl_y}(\alpha)\left[
Y_{l_x}(\widehat{x})\otimes Y_{l_y}(\widehat{y})\right]_{lm_l},\\
\Psi_K^{l_xl_y}(\alpha)&=&N_K^{l_xl_y}(\sin\alpha)^{l_x}(\cos
\alpha)^{l_y}\nonumber\\ &\times & P_n^{l_x+1/2,l_y+1/2} (\cos
2\alpha),
\end{eqnarray}
where $l_x$ and $l_y$ are the orbital angular momenta associated
with the Jacobi coordinates $x$ and $y$, respectively, $K$ is the
hypermomentum and $P_n^{a,b}$ is a Jacobi polynomial
with $n=(K-l_x-l_y)/2$. Usually the
hyperradial part is written as $R_{\beta}(\rho)=\rho^{-5/2}
U_{\beta} (\rho)$. Then, for bound states  $U(\rho)\to 0$ for
$\rho\to\infty$ with asymptotic behaviour $U(\rho) \to\exp{(-k
\rho)}$ with $\hbar^2 k^2/2m=-E$.

\subsection{The THO method}

The problem is now to solve the Eq.\ (\ref{hamil}) and find its
eigenvalues and eigenvectors. For that purpose, one needs a basis
in which the Hamiltonian can be diagonalized. As mentioned above,
one can expand the channel wave function $\Psi_{\beta JM} (\rho,
\Omega)$ in the HH basis as
\begin{equation}
\Psi_{\beta JM}(\rho,\Omega)=R_{\beta}(\rho) \sum_{m_l
\sigma}\langle lm_lS\sigma|JM\rangle \Upsilon_{Klm_l
}^{l_xl_y}(\Omega)~X_{S}^{\sigma}. \label{basecanal}
\end{equation}
$R_{\beta}(\rho)$ is the hyperradial wave function and here we
will use the THO method to obtain it. The basic idea of the THO
method is to perform a local scale transformation (LST) of the
ground state wave function of the system under study into the
harmonic oscillator ground state wave function. We now obtain such
a LST for a three-body system. The ground state of the system
$\Psi_B(\rho,\Omega)$ is written as a linear combination of the
basis functions, Eq. (\ref{basecanal}),
\begin{equation}
\Psi_B(\rho,\Omega)=\sum_{\beta}R_{B\beta}(\rho)
\sum_{m_l\sigma}\langle lm_lS\sigma|J_B\nu\rangle
\Upsilon_{Klm_l}^{l_xl_y}(\Omega)~X_{S}^{\sigma}.
\end{equation}
Consequently, we will obtain a LST for each channel $\beta$
included in the ground state. This is done by transforming the
hyperradial wave function for each channel included in the bound
ground state function, $R_{B\beta}(\rho)$, into the corresponding
(same $K$) harmonic oscillator ground state wave function for that
channel $R_{0K}^{HO}(s)$. In the six-dimensional case, these
functions are
\begin{equation}
R_{iK}^{HO}(s)=N_{iK}s^KL_i^{K+2}(s^2)\exp{(-s^2/2)}
\end{equation}
where $N_{iK}$ is the normalization constant
\begin{equation}
N_{iK}=\sqrt{\frac{2\Gamma(i+1)}{\Gamma(i+K+3)}}.
\end{equation}
Here $L_i^{{K+2}}(x)$ is a generalized Laguerre polynomial with
$i$ the order of the polynomial and which describes the
hyperradial excitation. The LST for a given channel $\beta$
included in the bound ground state is defined by the equation
\begin{equation}
\int_0^{\rho_{\beta}}d\rho~\rho^5|R_{B\beta}(\rho)|^2=\int_0^{s}
ds^\prime~ s^{\prime 5}~|R^{HO}_{0K}~(s^\prime)|^2. \label{etleR}
\end{equation}
Once the LST for each channel $\rho_{\beta}(s)$ is obtained, the
THO basis is defined by applying the inverse transformation
$s_{\beta} (\rho)$ to the HO wave functions generated from the
corresponding ground state wave function,
\begin{equation}
R^{THO}_{i\beta}(\rho)=\frac{N_{iK}}{N_{0K}}R_{B\beta}(\rho)
L_i^{K+2}(s_{\beta}(\rho)^2).
\end{equation}
Usually, instead of $R^{THO}_{i\beta}(\rho)$, the following
hyperradial wave function is introduced
\begin{equation}
U_{i\beta}^{THO}(\rho)=\rho^{5/2}R_{i\beta}^{THO}(\rho)
\end{equation}
that fulfils the orthonormality relationship
\begin{equation}
\int_0^{\infty}d\rho~U^{THO}_{i\beta}(\rho)U^{THO}_{i'\beta}(\rho) =
\delta_{ii'}.
\end{equation}

Thus, we obtain the THO basis
\begin{eqnarray}
\Psi_{i\beta JM}^{THO}(\rho,\Omega)&=&\rho^{-5/2}~U^{THO}_{i\beta}(\rho)\\
&\times&\sum_{m_l\sigma}\langle lm_lS\sigma|JM\rangle \Upsilon_{Klm_l}^{l_x
l_y}(\Omega)~X_{S}^{\sigma}.
\label{btho}
\end{eqnarray}
It is easy to show that these states form a complete orthonormal
set.

Since the THO basis is infinite, one diagonalizes the Hamiltonian
in a finite truncation, obtaining eigenstates
\begin{equation}
\Psi_{nJM}(\rho,\Omega)=\sum_{i\beta}C_{nJM}^{i\beta}\Psi_{i\beta
JM}^{THO}(\rho,\Omega)
\end{equation}
Convergence of the results with the basis truncation must then be
checked.

The information available on the different channels included in
the ground state wave function allows one to construct the
corresponding LST's directly, Eq. (\ref{etleR}). For channels not
included in the ground state, as a general rule, information from
one of the known (ground state) channels with the closest quantum
labels to the channel of interest is used to construct the LST.
One important point concerns the label $K$ which governs the
$\rho^K$ behavior of the hyperradial wave function close to the
origin. In order to keep this behavior correct we always select a
channel from the ground state wave function with the same $K$ as
the channel under study. If this is not possible, a channel with
$K-1$ is used and the corresponding hyperradial wave function is
then multiplied by $\rho$.

So, for example, the hyperradial wave function for
$\beta=\{1,1,0,1,1\},~J^{\pi}=1^-$ is obtained by multiplying by
$\rho$ the wave function for
$\beta=\{0,0,0,0,0\},~J^{\pi}=0^+$. The wave function for
$\beta=\{2,2,0,2,0\},~J^{\pi}=2^+$ is taken to be the same as the
one for
$\beta=\{2,0,0,0,0\},~J^{\pi}=0^+$. It is worth noting that
$l_x+l_y\le K$, while for the ground state components $l_x=l_y$.

\subsection{Hamiltonian matrix elements}

Recalling that one set of Jacobi coordinates has been selected
$\{\vec{x},\vec{y},\vec{R}\}$, and removing the CM contribution,
the Hamiltonian is written in hyperspherical coordinates as
\begin{equation}
\widehat{H}(\rho,\Omega)=\widehat{T}(\rho,\Omega)+\widehat{V}(\rho,\Omega)
\end{equation}
The kinetic energy operator, in the differential equations for the
$U(\rho)=\rho^{5/2}R(\rho)$, is
\begin{equation}
\widehat{T}(\rho,\Omega)=-\frac{\hbar^2}{2m}\left[ \frac{\partial^2}
{\partial\rho^2}+\frac{5}{\rho}\frac{\partial}{\partial\rho}-\frac{1}
{\rho^2}\widehat{K}^2(\Omega)  \right].
\end{equation}
The kinetic energy operator does not connect channels with
different $\beta$ or $J$, i.e.
\begin{eqnarray}
&\langle THO;i,\beta,J|\widehat{T}|THO;i',\beta',J'\rangle=
\nonumber\\ &\langle
THO;i,\beta,J|\widehat{T}|THO;i',\beta,J\rangle
\delta_{\beta,\beta'}\delta_{J,J'},
\end{eqnarray}
and hence only matrix elements within these channels have to be
calculated. These matrix elements are, with $s_\beta^2\equiv
s_\beta(\rho)^2$,
\begin{widetext}
\begin{eqnarray}
\langle THO;i,\beta,J|\widehat{T}|THO;i',\beta,J\rangle &=&
\frac{N_{iK}N_{i'K}}{N_{0K}^2}\left[ \int_0^{\infty}d\rho~
\frac{dU_{0\beta}(\rho)}{d\rho}^2L_{i}^{K+2}(s_\beta^2)L_{i'}^{K+2}
(s_\beta^2)\right. \nonumber \\
&+& \int_0^{\infty}d\rho~ U_{0\beta
}(\rho)^2L_{i-1}^{K+3}(s_\beta^2)L_{i'-1}^{K+3}(s_\beta^2)\left(2s_{\beta
}(\rho)\frac{ds_{\beta }(\rho)}{d\rho}\right)^2
\nonumber \\
&-& \int_0^{\infty}d\rho~ \frac{dU_{0\beta
}(\rho)}{d\rho}U_{0\beta
}(\rho)L_{i-1}^{K+3}(s_\beta^2)L_{i'}^{K+2}(s_\beta^2)2s_{\beta
}(\rho)\frac{ds_{\beta }(\rho)}{d\rho}
\nonumber \\
&-& \int_0^{\infty}d\rho~ \frac{dU_{0\beta
}(\rho)}{d\rho}U_{0\beta
}(\rho)L_{i}^{K+2}(s_\beta^2)L_{i'-1}^{K+3}(s_\beta^2)2s_{\beta
}(\rho)\frac{ds_{\beta }(\rho)}{d\rho}
\nonumber \\
&+& \left.\left(\frac{15}{4}+K(K+4)\right)\int_0^{\infty}d\rho~
\frac{U_{0\beta
j}(\rho)^2}{\rho^2}L_{i}^{K+2}(s_\beta^2)L_{i'}^{K+2}(s_\beta^2)
\rule[0cm]{0cm}{0.7cm}\right].
\end{eqnarray}
\end{widetext}
These integrals are evaluated by quadratures.

The potential energy operator connects different channels within
the same $J$. The hyperangular integration is performed using the
code FaCE \cite{face} on the set of hyperradial quadrature points
providing the set of functions $V_{\beta\beta'}^J(\rho)$. The
potential energy matrix elements in the THO basis are then simply
\begin{widetext}
\begin{equation}
\langle THO;i,\beta,J|\widehat{V}(\rho,\Omega)|THO;i',\beta',J\rangle=
\int_0^{\infty}d\rho~ U_{i\beta }^{THO}(\rho)V_{\beta\beta'}^J(\rho)
U_{i'\beta }^{THO}(\rho).
\end{equation}
\end{widetext}

The formalism presented above provides a complete basis and the
corresponding Hamiltonian matrix elements. In the following
Section we apply the method to study $^6$He.

\section{Structure calculation for $^6$He}

The $^6$He nucleus is treated here as a three-body system,
comprising an inert $\alpha$ core and two valence neutrons. The
ground state has total angular momentum $J^{\pi}=0^+$ and an
experimental binding energy of $0.973$ MeV.  The ground state wave
function was obtained by solving the Schr\"odinger equation in
hyperspherical coordinates, following the procedure described in
\cite{Zhu93,face}. Besides the two-body ($n-n$ and $n-\alpha$)
potentials, the model Hamiltonian also includes a simple central
hyperradial three-body force. This is introduced to overcome the
under-binding caused by the other closed channels, the most
important of which are $t$+$t$ channels. The $n$-$^4$He potential
is taken from Ref.~\cite{Bang79,Tho00}, with central and
spin-orbit components, and the GPT $NN$ potential \cite{gpt} with
central, spin-orbit and tensor components is used. These
calculations were performed with the code FaCE \cite{face}. All
calculations truncate the maximum hypermomentum at $K_{max}=20$
and the three-body force is adjusted to give the right binding
energy. The calculated three-body wave function has a binding
energy of $0.954882$ MeV and a rms point nucleon matter radius of
$2.557$ fm when assuming an alpha-particle rms matter radius of
$1.47$ fm.

\begin{figure}
\resizebox*{0.5\textwidth}{!}{\includegraphics{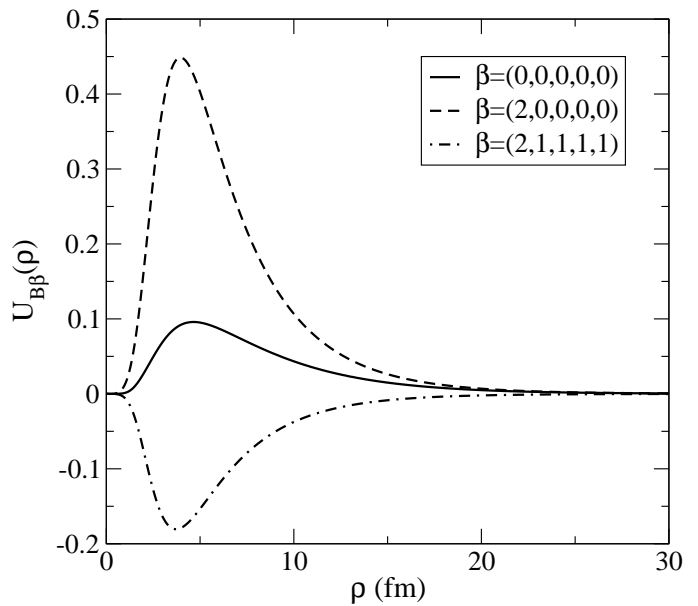}}
\vspace{-0.4cm}
\caption{\label{wfgs} Radial part for the dominant components of the $^6$He ground
state wavefunction. The labels stand for the angular momentum quantum numbers,
in the order  $(K,l_x,l_y,l, S)$}
\end{figure}

In Fig.~\ref{wfgs} we plot the hyperradial parts for the first
three channels of the $^6$He ground state wave function.  The
labels correspond to the quantum numbers $\{K,l_x,l_y,l, S\}$.
These channels give the main contribution to the wave function.

\begin{figure}
\resizebox*{0.5\textwidth}{!}{\includegraphics{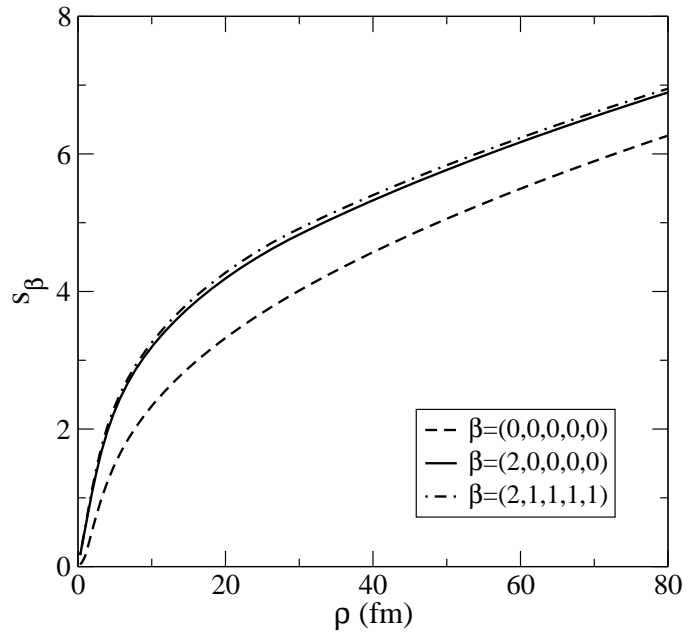}}
\vspace{-0.4cm} \caption{\label{LSTs}Local scale transformations
for the first three channels included in $^6$He ground state
wavefunction.}
\end{figure}

From the ground state, the LST for the different channels are
obtained as explained in the preceding section. In Fig.~\ref{LSTs}
the $s_{\beta}(\rho)$ for the LST's of the most important $^6$He
ground state channels are shown. The THO basis is constructed from
these LST using Eq.~(\ref{btho}).

\begin{figure}
\resizebox*{0.5\textwidth}{!}{\includegraphics{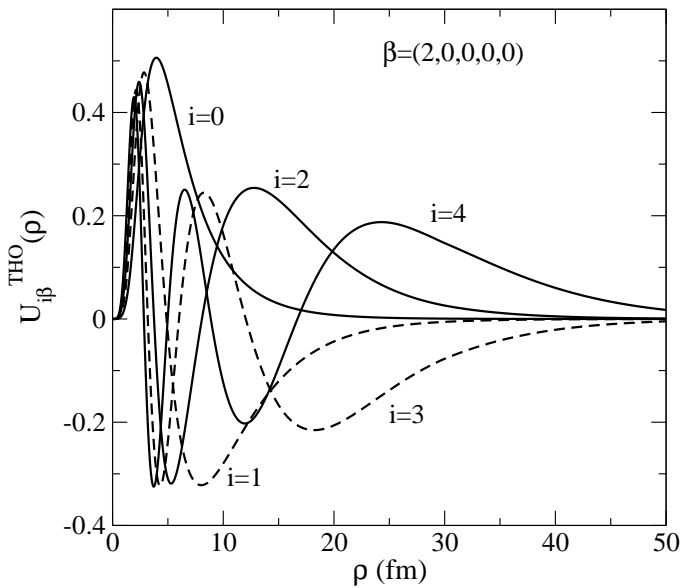}}
\vspace{-0.4cm} \caption{\label{thowf} First five THO states for
the $J=0$ channel $\beta=\{2,0,0,0,0\}$. }
\end{figure}

In Fig.~\ref{thowf} the first few hyperradial wave functions for
the channel $\beta=\{2,0,0,0,0\}$, and for $J^{\pi}=0^+$, are
presented. This is the most important ground state channel, making
a 79\% contribution to the total norm. We see that as the quantum
number $i$ increases the wave functions are more oscillatory and
explore larger distances.

\begin{figure}
\resizebox*{0.45\textwidth}{!}{\includegraphics{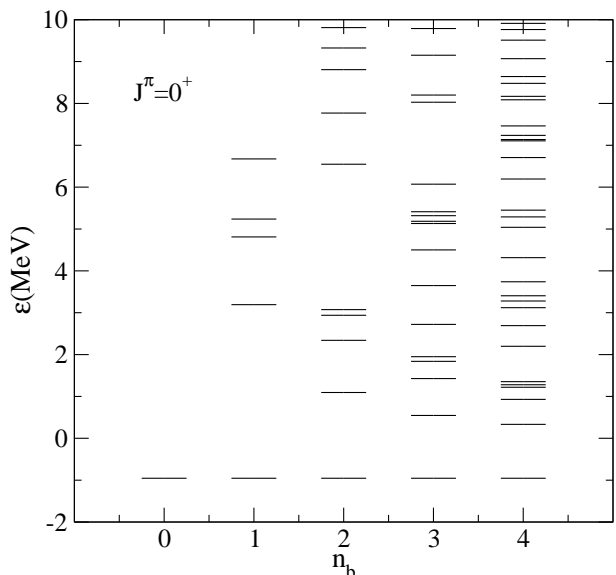}}
\vspace{-0.4cm} \caption{\label{thoener} Eigenvalues of
$J^{\pi}=0^+$ states below $\varepsilon=10$MeV. }
\end{figure}

With the THO basis one can proceed to calculate the Hamiltonian
matrix elements and diagonalize the Hamiltonian matrix in a
truncated space. In Fig.~\ref{thoener} the Hamiltonian eigenvalues
for $J^{\pi}=0^+$, for different maximum values of the hyperradial
excitations, $n_b$, are presented up to $\varepsilon=10$ MeV. The
calculated ground state energy is $-0.954886$ MeV.

To study different strength functions from the $^6$He ground state
to the three-body continuum, states with different values of
$J^{\pi}$ have to be generated. Thus, we have to construct a THO
basis for these states starting from the information we have for
the ground state. Close to the origin the hyperradial wave
functions behave as $\rho^K$. Thus, in order to calculate the LST
for states with $J=2$ and hypermomentum $K$ ($K=2,4,6,\dots,2n$)
we use the ground state component with $J=0$ and the same value
for $K$ (and also $l_x,l_y,l,S$ if needed). Again, using the LST
we generate the basis for $J=2$ and diagonalize the Hamiltonian.

\begin{figure}
\resizebox*{0.45\textwidth}{!}{\includegraphics{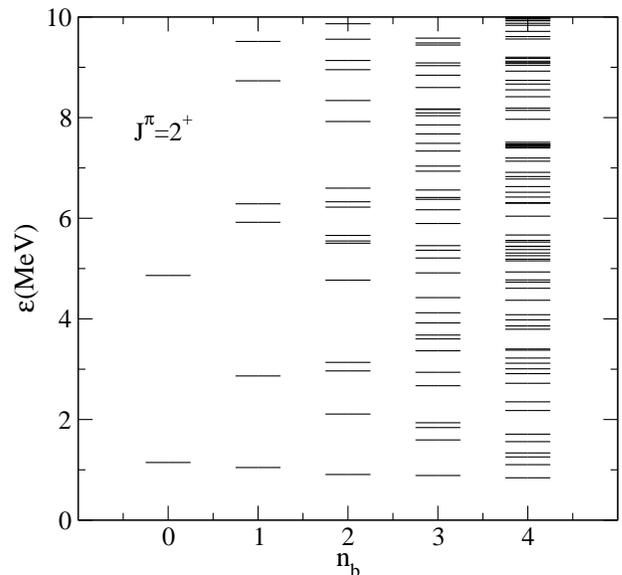}}
\vspace{-0.4cm} \caption{\label{thoener2} Eigenvalues of
$J^{\pi}=2^+$ states below $\varepsilon=10$MeV. }
\end{figure}

In Fig.~\ref{thoener2}, the eigenvalues of the Hamiltonian for
$J^\pi=2^+$ states are presented for different values of $n_b$.
Again the lowest state is very stable and is close to the energy
of the known $2^+$ resonance, that is $0.824$ MeV. The states with
$J=1$ have odd $K$ ($K=1,3,5,...,2n-1$). Since the ground state
wave function contains only even $K$, for the generation of the
LST for the $J=1$ state we have taken the ground state component
with $K-1$ and multiplied this by $\rho$, to recover the correct
behaviour close to the origin. To select different channels with
the same value for K we look for coincidence in $l_x,l_y,l,S$.
Again, from the LST we generate the basis for $J=1$ and
diagonalize the Hamiltonian.

\begin{figure}
\resizebox*{0.45\textwidth}{!}{\includegraphics{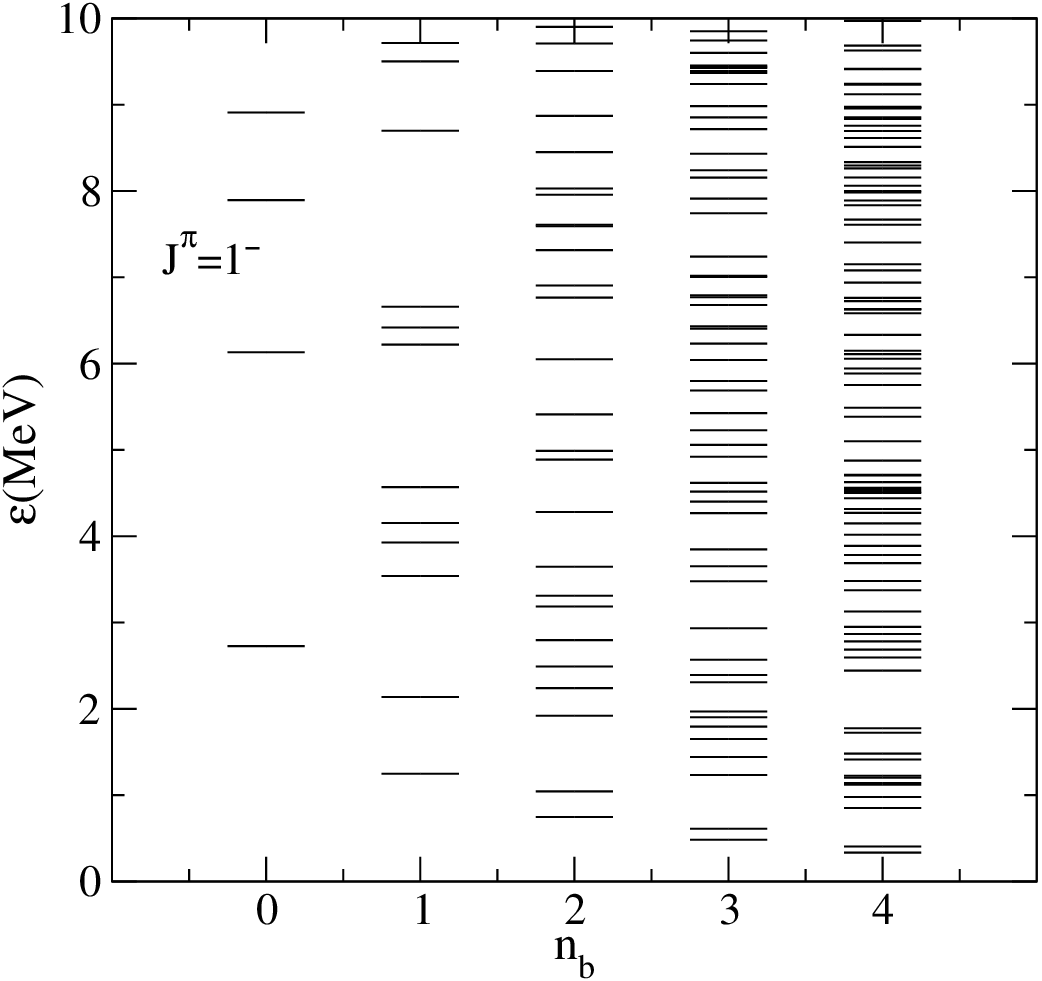}}
\vspace{-0.4cm} \caption{\label{thoener1} Eigenvalues of $J=1^-$
states below $\varepsilon=10MeV$.  }
\end{figure}

In Fig.~\ref{thoener1}, the eigenvalues of the Hamiltonian for
$J^\pi=1^-$ states are presented for different values of $n_b$. We
now see that the lowest state changes as a function of $n_b$, and
goes down in energy as the dimension of the basis is increased.

We now calculate response functions of our system and assess their
convergence with the dimension of our truncation of the THO basis.

\subsection{Completeness}
\label{complet}

In general, a pseudostate with energy $\varepsilon_n$ will be a
superposition of the actual continuum states nearby in energy.
There are different ways of assigning an energy distribution to a
pseudostate \cite{Mac87a,Bra95,Cre02}. Here we propose a method
that takes as reference a large basis with $N_t=(N_b+1)\times
N_{chan}$ states, where $N_b$ is the number of hyperradial
excitations of the large basis and $N_{chan}$ is the number of
channels for a given $J^{\pi}$. This basis is considered to be
complete for the problem under study. In this basis an energy
distribution $f_N(\varepsilon,\varepsilon_N)$ is assigned to each
discrete state ($N=1,\ldots,N_t$). The width of the distribution
is an increasing function of the energy and the distribution can
be Gaussian, Lorentzian, or Poisson, etc. We then consider a
smaller basis, with $n_t\ll N_t$ states, $n_t=(n_b+1)\times
N_{chan}$, where $n_b$ is the number of hyperradial excitations of
the smaller basis. These states $n$ ($n=1,\ldots,n_t$) can be
expanded in the large basis
\begin{equation}
|n\rangle=\sum_{N}C(n,N)|N\rangle.
\end{equation}
Then, the distribution for the states in the small basis $f_n
(\varepsilon,\varepsilon_n)$ is
\begin{eqnarray}
f_n(\varepsilon, \varepsilon_n)&=&|\langle
\varepsilon|n\rangle|^2=\sum_{NN'}\langle N|\varepsilon\rangle
\langle
\varepsilon |N' \rangle C(n,N)C(n,N')^* \nonumber \\
&\approx&
\sum_{N}|\langle \varepsilon|N\rangle|^2|C(n,N)|^2 \nonumber\\
&= &\sum_{N}f_N(\varepsilon,\varepsilon_N)|C(n,N)|^2.
\end{eqnarray}
where we have considered that the off-diagonal terms are small
compared with the diagonal terms in the bigger basis.

We have approximated the $f_N(\varepsilon,\varepsilon_N)$ by a
Poisson distribution
\begin{equation}
P(\varepsilon,\varepsilon_N,m)=\frac{(m+1)^{(m+1)}}{
\varepsilon_N^{(m+1)}m!}~\varepsilon^m\exp\left(-\frac{m+1}
{\varepsilon_N}~ \varepsilon\right),
\end{equation}
which has a width given by
\begin{equation}
\Gamma_N=\sqrt{\langle \varepsilon^2\rangle-\langle
\varepsilon\rangle^2}=\frac{\varepsilon_N}{\sqrt{m+1}}.
\end{equation}
The larger the value selected for $m$ the narrower the distribution is.

\begin{figure}
\resizebox*{0.45\textwidth}{!}{\includegraphics{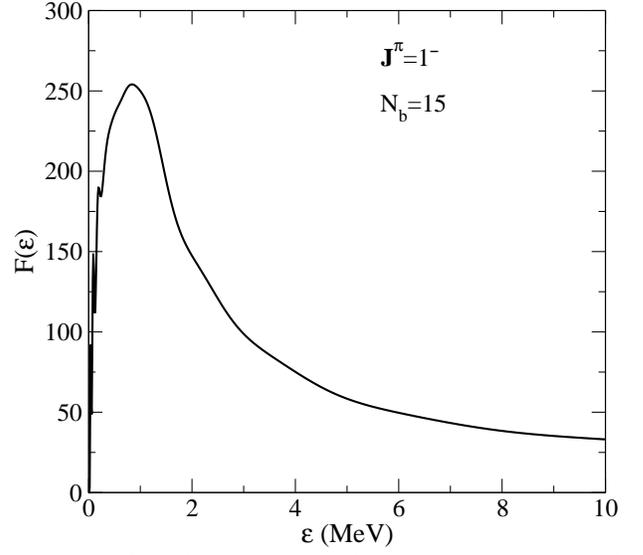}}
\vspace{-0.4cm} \caption{\label{comple1} Completeness of the basis
with $N_b=15$ and a Poisson distribution with $m=20$ for the
$J^{\pi}=1^-$ states.}
\end{figure}

\begin{figure}
\resizebox*{0.45\textwidth}{!}{\includegraphics{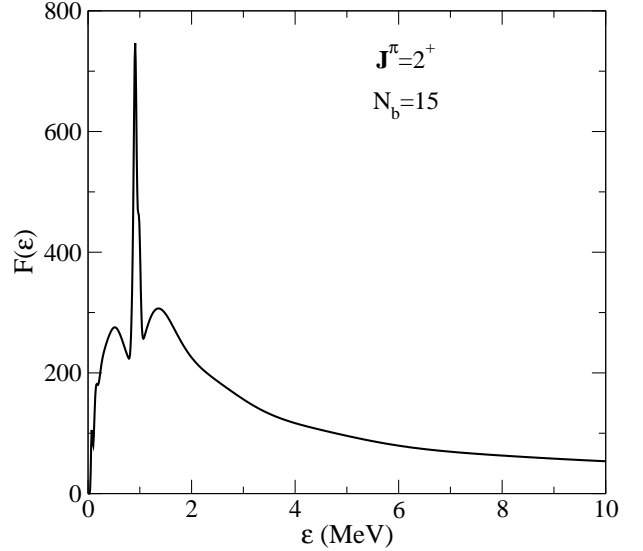}}
\vspace{-0.4cm} \caption{\label{comple2} Completeness of the basis
with $N_b=15$ and a Poisson distribution for the $J=2^+$ states.
See text for details on the $m$ values used.}
\end{figure}

In Figs.~\ref{comple1} and \ref{comple2} we present the
completeness of the large basis to describe states with
$J^{\pi}=1^-$ and $J^{\pi}=2^+$ respectively. The completeness
$F(\varepsilon)$ is defined as the sum over all the excited states
of $P(\varepsilon,\varepsilon_N,m)$. For the $B(E1)$ distribution
a value of $m=20$ was used. For the $B(E2)$ the same value of $m$
was used except for the region around the resonance for which a
value $m=1300$ had to be used in order to get a width consistent
with the experimental value. It is observed that the basis is
particularly suited for describing effects in the low lying
continuum in both cases.

\subsection{Strength functions}

To study the effectiveness of the THO basis we start with global
observables related to structure, the strength functions. For a
given operator $\widehat{O}$, which couples the ground state
($n=1,J^{\pi}=0^+$) with excited states with angular momentum
$J^{\pi}$, the following observables are defined:
\begin{itemize}
\item Total strength
\begin{equation}
S_T(\widehat{O},n_t)=\sum_n^{n_t}|\langle nJ|\widehat{O}|10\rangle |^2
\end{equation}
where $n$ runs over all the excited states with total angular
momentum  $J^{\pi}$, up to $n_t=(n_b+1)\times N_{chan}$.

In the limit of a very large number of states $n_t$, the sum rule
for the total strength is
\begin{eqnarray}
S_T(\widehat{O})&=&\lim_{n_t\to\infty}S_T(\widehat{O},n_t) \nonumber\\
&=&\langle 10|\widehat{O}^2|10\rangle -|\langle 10| \widehat{O}|
10\rangle |^2.
\end{eqnarray}
\item Energy weighted sum
\begin{equation}
E_w(\widehat{O},n_t)=\sum_n^{n_t}(\varepsilon_{nJ}-\varepsilon_{10})
|\langle nJ|\widehat{O}|10\rangle |^2.
\end{equation}
\item Polarizability
\begin{equation}
\alpha(\widehat{O},n_t)=\sum_n^{n_t}(\varepsilon_{nJ}-\varepsilon_{10}
)^{-1}|\langle nJ|\widehat{O}|10\rangle |^2.
\end{equation}
\end{itemize}

In Table \ref{T1} we present the results for the case
$\widehat{O}=\rho^2$ that connects the ground state with the
excited states with $J^{\pi}=0^+$. This calculation is performed
with $K_{max}=20$ and different values of $n_b$.
\begin{table}
\caption{Convergence of different observables as the THO basis is
increased: calculated ground state energy ($e_B$), expectation
value in the ground state for $\rho^2$ ($\langle\rho^2\rangle_B$),
total strength from the ground state ($S_T$) [this column should
converge to the column labelled $\langle\rho^4 \rangle_B- \langle
\rho^2\rangle_B^2$ according to the sum rule], the ratio between
the energy weighted sum and the total strength ($\langle E(n_t)
\rangle$) and the ratio between the polarizability and the total
strength ($\langle E^{-1}(n_t)\rangle$).}

\begin{ruledtabular}
\begin{tabular}{cccccccc}
$n_b$ & $e_B$ & $\langle\rho^2\rangle_B$ & $S_T$ & $\langle\rho^4
\rangle_B-\langle\rho^2\rangle_B^2$ & $\langle E\rangle$ & $\langle
E^{-1}\rangle$ \\
 & (MeV) & (fm$^2$) & (fm$^4$) & (fm$^4$) & (MeV) & (MeV$^{-1}$) \\
\colrule
$0$ & $-0.954882$ & $30.63$ & $41.6$ &  $855$ & $27.8$ & $0.0433$\\
$1$ & $-0.954884$ & $30.60$ & $738$ &  $853$ & $5.72$ & $0.194$\\
$2$ & $-0.954885$ & $30.62$ & $851$ &  $856$ & $3.08$ & $0.364$\\
$3$ & $-0.954885$ & $30.62$ & $855$ &  $855$ & $3.01$ & $0.426$\\
$4$ & $-0.954886$ & $30.62$ & $854$ &  $854$ & $3.00$ & $0.433$ \\
\end{tabular}
\end{ruledtabular}
\label{T1}
\end{table}

Note that, already for $n_b=2$, the total strength acquires a
value which is very close to the sum rule value. This indicates
that the THO basis is very efficient to describe the ground state
through the relatively long range monopole operator $\rho^2$. The
values of $\langle E(n_t)\rangle$ and $\langle E^{-1}(n_t)\rangle$
also stabilize for low values of $n_b$ and indicate the range of
energies in the continuum which are relevant. In our case, for
$J^{\pi}=0^+$, it corresponds to low energies of around $3$ MeV
excitation energy above the break-up threshold.

\begin{table}
\caption{\label{SRE1} Convergence of observables as a function of
the dimension ($n_b$) of the basis considered. The observables
presented are: the total strength for B(E1) excitations from the
ground state ($\sum_nB(E1)_{10,n1}$), the energy weighted sum
($E_w$), the ratio between these magnitudes ($\langle E\rangle
={E_w}/{\sum_nB(E1)_{10,n1}}$) and the polarizability ($\alpha$)
below $\varepsilon=10$MeV.}
\begin{ruledtabular}
\begin{tabular}{ccccc}
$n_b$ & $\sum_nB(E1)_{10,n1}$ & $E_w$ & $\langle E\rangle$ &
$\alpha$\\ & (e$^2$fm$^2)   $          & (e$^2$fm$^2$MeV)  &
(MeV)             & (e$^2$fm$^2$MeV$^{-1}$)\\
\colrule
$0$ & $1.297$ & $5.38$ & $4.15$ & $0.422$\\
$1$ & $1.323$ & $5.71$ & $4.32$ & $0.403$\\
$2$ & $1.323$ & $5.76$ & $4.36$ & $0.403$\\
$3$ & $1.321$ & $5.76$ & $4.36$ & $0.402$\\
$4$ & $1.320$ & $5.75$ & $4.35$ & $0.402$\\
\end{tabular}
\end{ruledtabular}
\end{table}

\begin{table}
\caption{\label{SRE2} Same as Table \ref{SRE1} but for E2 transitions.}
\begin{ruledtabular}
\begin{tabular}{ccccc}
$n_b$ & $\sum_nB(E2)_{10,n2}$ & $E_w$          & $\langle E\rangle$
& $\alpha$\\ & (e$^2$fm$^4)   $          & (e$^2$fm$^4$MeV)  &
(MeV)             & (e$^2$fm$^4$MeV$^{-1}$)\\

\colrule
$0$ & $3.51$ & $14.5$ & $4.09$ & $1.21$\\
$1$ & $6.89$ & $27.6$ & $4.01$ & $2.37$\\
$2$ & $8.41$ & $31.1$ & $3.70$ & $3.06$\\
$3$ & $8.47$ & $30.5$ & $3.60$ & $3.20$\\
$4$ & $8.41$ & $30.1$ & $3.58$ & $3.19$\\

\end{tabular}
\end{ruledtabular}
\end{table}

\subsection{B(E$\lambda$) sum rules}

We will follow the notations and definitions of Brink and Satchler
\cite{BS94}. The electric multipole operator $Q_{\lambda M_{
\lambda }}(\vec{r})$ is defined as
\begin{equation}
Q_{\lambda M_{\lambda}}(\vec{r})=\left(\frac{4\pi}{2\lambda+1}
\right)^{1/2}Z~e~r^{\lambda}Y_{\lambda M_{\lambda}}(\widehat{r}),
\end{equation}
where $e$ is the electron charge and $Z$ is the atomic number of
the system. For a nucleus with a core plus two valence neutrons,
$\vec{r}=\vec{y}\sqrt{m\mu_y}/m_c$ is the position of the core
(charged particle) relative to the center of mass of the system.
$m_c$ is the mass of the core. For $^6$He, $Z=2$ and $\vec{r}
=\vec{y}/(2\sqrt{3})$, as presented in Fig. \ref{jacobi-6he}. The
reduced transition probability is
\begin{eqnarray}
B(E\lambda)_{nJ,n'J'} & \equiv & B(E\lambda;\, nJ \to n'J')\nonumber\\
&=&|\langle nJ||Q_{\lambda}||n'J'\rangle
|^2\left(\frac{2\lambda+1}{4\pi}\right)
\end{eqnarray}
where the reduced matrix element $\langle n J||Q_{\lambda}||n'J'
\rangle$ is defined as
\begin{eqnarray}
\langle n JM|Q_{\lambda
M_{\lambda}}|n'J'M'\rangle&=&(-1)^{2\lambda}\langle J'M'\lambda
M_{\lambda}|JM \rangle \nonumber\\ &\times& \langle
nJ||Q_{\lambda}||n'J'\rangle . \label{emr}
\end{eqnarray}
Using some angular momentum algebra we arrive to the final expression
\begin{widetext}
\begin{eqnarray}
B(E\lambda)_{nJ,n'J'}&=&\frac{2\lambda+1}{4\pi}|\langle
nJ||Q_{\lambda}||n'J'\rangle|^2 \nonumber \\
&=&\frac{(2\lambda+1)(2J'+1)}{4\pi}Z^2~e^2\left(\frac{\sqrt{m
\mu_y}}{m_c}\right)^{2\lambda}
\nonumber \\
&\times& \left|\sum_{\beta\beta'}\delta_{SS'}\delta_{l_xl'_x}
\sum_{ii'}C_{n}^{i\beta}C_{n'}^{i'\beta'}
\hat{l}\hat{l}'\hat{l}_y\hat{l}'_y(-)^{l_x+S} \right.  \nonumber \\
&\times&
\left(\begin{array}{ccccc}l_{y}~\lambda~l'_{y}\\0~0~0\end{array}\right)
W(ll'l_yl'_y;\lambda l_x)W(JJ'll';\lambda S) \nonumber \\
&\times& \left. \int\int(\sin{\alpha})^2(\cos{\alpha})^2d\alpha~
d\rho~ y^{\lambda}
U_{i\beta}^{THO}(\rho)\Psi_{K}^{l_{x}l_{y}}(\alpha)
\Psi_{K'}^{l_{x}l'_{y}}(\alpha)U_{i'\beta'}^{THO}(\rho)\right|^2.
\end{eqnarray}
\end{widetext}

Using this expression it is simple to calculate the sum rules of
electric transitions from the ground state, $n=1$, $J^{\pi}=0^+$,
to the states $(n,J)$: $\sum_nB(E\lambda)_{10,nJ}$,
\begin{equation}
\sum_nB(E\lambda)_{10,nJ}=\left(\frac{2\lambda+1}{4\pi}\right)
\sum_n|\langle 1~0||Q_{\lambda}||nJ\rangle |^2.
\end{equation}
In particular, we are interested in the $\lambda=1$ electric
dipole strength, connecting the $J^{\pi}=0^+$ ground state to the
$J^{\pi}=1^-$ states, and in the $\lambda=2$ electric quadrupole
strength, connecting the ground state to the $J^{\pi}=2^+$ states:
\begin{itemize}
\item B(E1) sum rule
\begin{equation}
\sum_nB(E1)_{10,n1}=\frac{3}{4\pi}\frac{Z^2e^2m\mu_y}{m_c^2}
\langle 1~0|y^2|1~0\rangle.
\end{equation}

\item B(E2) sum rule
\begin{equation}
\sum_nB(E2)_{10,n2}=\frac{5}{4\pi}\frac{Z^2e^2m^2\mu_y^2}{m_c^4}
\langle 1~0|y^4|1~0\rangle.
\end{equation}
\end{itemize}

Tables \ref{SRE1} and \ref{SRE2} show the results obtained for
$^6$He when including states up to 10 MeV in excitation energy
above the two-neutron break-up threshold. These include the
$B(E\lambda)$, total strength, energy weighted sum, and
polarizability. The $B(E1)$ and $B(E2)$ sum rule values for the
total strength are  $1.500$ e$^2$fm$^2$ and $9.99$ e$^2$fm$^4$,
respectively. The values in the second column in Tables \ref{SRE1}
and \ref{SRE2} are close to these limits but do not reach them
since only states up to $10$MeV are included. If the complete sum
is done the values are $1.498$ e$^2$fm$^2$ and $9.59$ e$^2$fm$^4$
that are close to the corresponding sum rule values.

\subsection{B(E1) and B(E2) distributions}

To obtain a continuous $B(E\lambda)$ distribution from the
discrete values $B(E\lambda)_{10,nJ}$ we can apply the procedure
described in subsection \ref{complet}. In this case we have to
evaluate
\begin{widetext}
\begin{eqnarray}
|\langle gs|\widehat{O}|\varepsilon \rangle|^2&=&\langle gs
|\widehat{O}|\varepsilon\rangle\langle
\varepsilon|\widehat{O}|gs\rangle\approx \sum_{nn'}
\langle gs|\widehat{O}|n\rangle\langle
n|\varepsilon\rangle\langle \varepsilon|n'\rangle\langle
n'|\widehat{O}|gs\rangle \nonumber \\
&=& \sum_{nn'}\langle gs|\widehat{O}|n\rangle\langle
n'|\widehat{O}|gs\rangle\sum_{NN'}\langle N|\varepsilon
\rangle\langle \varepsilon|N'\rangle
C(n,N)C(n',N')^* \nonumber \\
&\approx&
\sum_{nn'}\langle gs|\widehat{O}|n\rangle\langle
n'|\widehat{O}|gs\rangle\sum_{N}C(n,N)f_N (\varepsilon,
\varepsilon_N) C(n',N)^* \nonumber \\
&=& \sum_{N}f_N(\varepsilon,\varepsilon_N)SO(N)SO(N)^*
\end{eqnarray}
\end{widetext}
where $SO(N)=\sum_{n}\langle gs|\widehat{O}|n\rangle C(n,N)$. We
have used this method to calculate $B(E\lambda)$ distributions.
For this purpose we have taken a large basis with $N_t=(N_b
+1)\times N_{chan}$ states, where $N_b=15$. The Poisson
distribution used had $m=20$.

\begin{figure}
\resizebox*{0.45\textwidth}{!}{\includegraphics{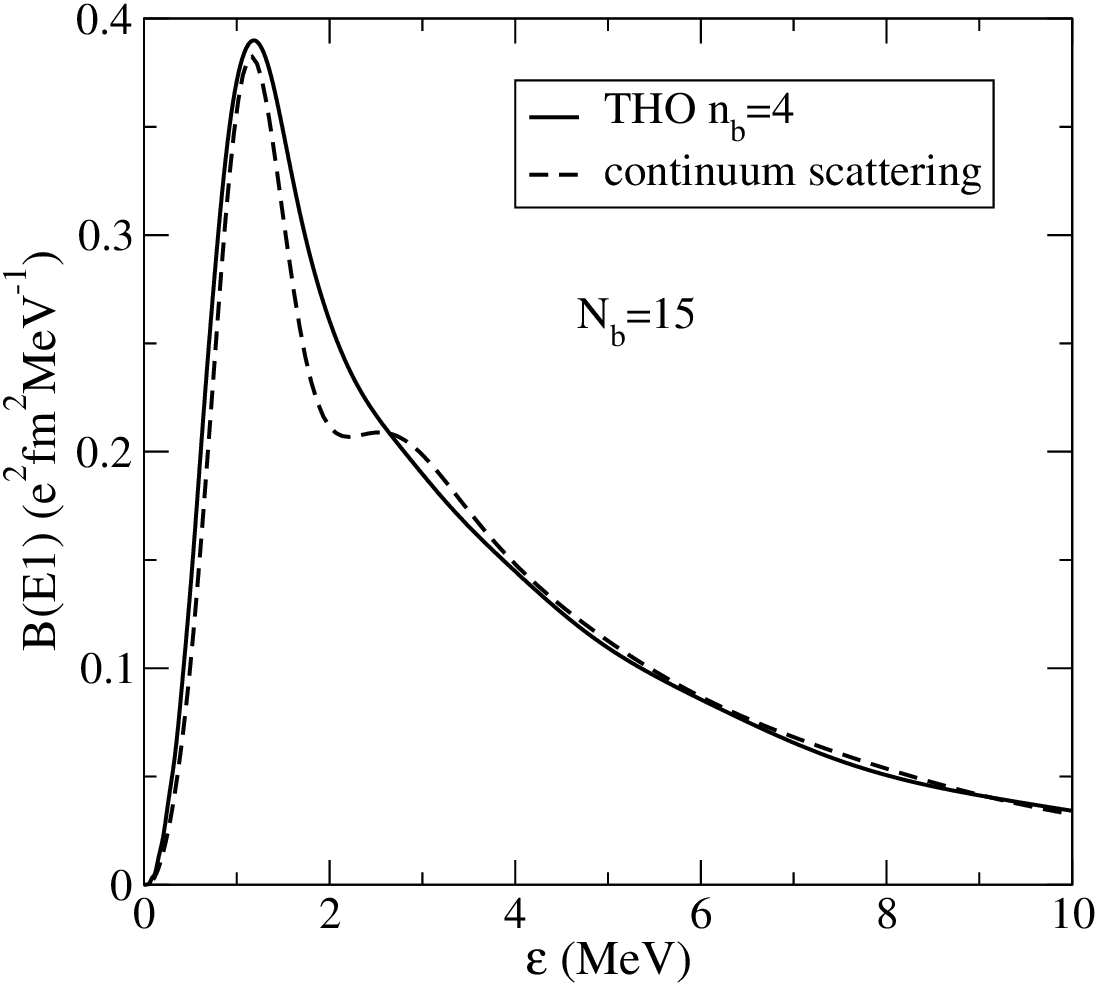}}
\vspace{-0.4cm} \caption{\label{BE1dis} $B(E1)$ distribution for
$n_b=4$ with $n_t=(n_b+1)\times N_{chan}$ and a Poisson
distribution with $m=20$.}
\end{figure}

\begin{figure}
\resizebox*{0.45\textwidth}{!}{\includegraphics{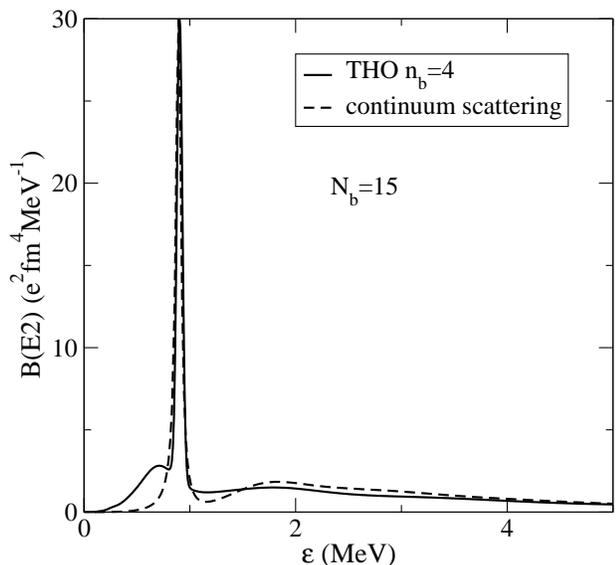}}
\vspace{-0.4cm} \caption{\label{BE2dis} $B(E2)$ distribution for
$n_b=4$ with $n_t=(n_b+1)\times N_{chan}$ and a Poisson
distribution with $m=20$.}
\end{figure}

\begin{figure}
\resizebox*{0.45\textwidth}{!}{\includegraphics{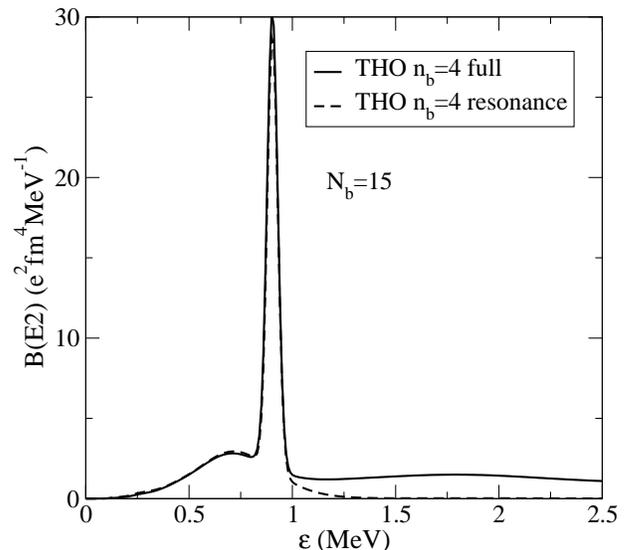}}
\vspace{-0.4cm} \caption{\label{BE2res}Total $B(E2)$ distribution
for $n_b=4$ and resonance $B(E2)$ distribution for $n_b=4$ and for
a Poisson distribution with $m=20$.}
\end{figure}

In Figs.~\ref{BE1dis} and \ref{BE2dis} we present the $B(E1)$ and
$B(E2)$ distributions up to 10 MeV. Calculations for different
dimensions of the THO basis are not presented, but are practically
indistinguishable. For comparison, the distribution calculated
with the continuum scattering wave functions is also shown. The
latter is the calculation reported in Ref.~\cite{Tho00}. For the
$B(E1)$ the distributions are in good agreement with the maximum
at around $1.2$ MeV and with the same total strength. For the
$B(E2)$ distributions the most relevant feature is the appearance
of a narrow low-lying resonance. In both calculations the depth of
the three-body interaction was adjusted to reproduce the position
of the $2^+$ resonance at $0.824$ MeV. Again we find good
agreement between the calculations in both shape and total
strength. As noted above, we used a larger $m$ value for the
energy region around the resonance. Our calculations exhibit a
small bump just below the resonance which does not appear in the
continuum scattering calculation. This small difference comes from
the low energy components of the resonance. This is more clearly
shown in Fig.~\ref{BE2res} where, besides the full $B(E2)$
distribution, we have superimposed the contribution of the first
$J^{\pi}=2^+$ state that appears around the resonance energy in
the THO calculation for $n_b=4$. This single state reproduces the
resonant peak very well.

\section{Summary and conclusions}
We have presented a generalization of the transformed harmonic
oscillator (THO) method proposed in \cite{Per01a} appropriate for
three-body problems. The method provides a discrete representation
of the continuum spectrum of a three-body system from a knowledge
of its ground state wave function, either in analytical or
numerical form. This wave function is used to obtain a local scale
transformation that converts this state into a harmonic oscillator
ground state. The inverse transformation is then applied to the
corresponding excited harmonic oscillator states to obtain a basis
set for the physical system. Finally, the three-body Hamiltonian
is diagonalized in this basis, providing a discrete representation
of eigenstates and eigenvalues of the three-body system.

The formalism has been applied here to the Borromean nucleus
$^6$He, for which several strength functions, including the dipole
and quadrupole Coulomb transition strengths, have been calculated.
These observables are found to converge quickly with respect to
the number of THO basis states included.  Furthermore, the
calculated strength distributions are in very good agreement with
previous results obtained using a three-body continuum scattering
wave functions. The results found in this work suggest that the
THO basis could also be effective for continuum discretization in
scattering calculations. Work in this direction is in progress and
results will be presented in a future publication.

\begin{acknowledgments}
This work was supported in part by the Spanish DGICYT under
project numbers BFM2002-03315 and FPA2002-04181-C04-04 and in part
by EPSRC under grant GR/M82141. A.M.M. acknowledges a research
grant from the Junta de Andaluc\'ia. M.R.G. acknowledges a
research grant from the Ministerio de Educaci\'on and the Marie
Curie Training Site. Enlightening discussions with Prof. R. C.
Johnson are deeply acknowledged.
\end{acknowledgments}


\bibliographystyle{unsrt}

\begin{thebibliography}{41}
\expandafter\ifx\csname natexlab\endcsname\relax\def\natexlab#1{#1}\fi
\expandafter\ifx\csname bibnamefont\endcsname\relax
  \def\bibnamefont#1{#1}\fi
\expandafter\ifx\csname bibfnamefont\endcsname\relax
  \def\bibfnamefont#1{#1}\fi
\expandafter\ifx\csname citenamefont\endcsname\relax
  \def\citenamefont#1{#1}\fi
\expandafter\ifx\csname url\endcsname\relax
  \def\url#1{\texttt{#1}}\fi
\expandafter\ifx\csname urlprefix\endcsname\relax\def\urlprefix{URL }\fi
\providecommand{\bibinfo}[2]{#2}
\providecommand{\eprint}[2][]{\url{#2}}

\bibitem[{\citenamefont{Hansen et~al.}(1995)\citenamefont{Hansen, Jensen, and
  Jonson}}]{Han95}
\bibinfo{author}{\bibfnamefont{P.~G.} \bibnamefont{Hansen}},
  \bibinfo{author}{\bibfnamefont{A.}~\bibnamefont{Jensen}}, \bibnamefont{and}
  \bibinfo{author}{\bibfnamefont{B.}~\bibnamefont{Jonson}},
  \bibinfo{journal}{Ann. Rev. Nucl. Part. Sci. {\bf 45 }, 591}
  (\bibinfo{year}{1995}).

\bibitem[{\citenamefont{Jensen et~al.}(2004)\citenamefont{Jensen, Riisager, and
  Garrido}}]{Jen04}
\bibinfo{author}{\bibfnamefont{A.~S.} \bibnamefont{Jensen}},
  \bibinfo{author}{\bibfnamefont{D.~V.} \bibnamefont{Riisager},
  \bibfnamefont{K.~Fedorov}}, \bibnamefont{and}
  \bibinfo{author}{\bibfnamefont{E.}~\bibnamefont{Garrido}},
  \bibinfo{journal}{Rev. Mod. Phys.} \textbf{\bibinfo{volume}{76}},
  \bibinfo{pages}{215} (\bibinfo{year}{2004}).

\bibitem[{\citenamefont{Yahiro et~al.}(1986)\citenamefont{Yahiro, Iseri,
  Kameyama, Kamimura, and Kawai}}]{Yah86}
\bibinfo{author}{\bibfnamefont{M.}~\bibnamefont{Yahiro}},
  \bibinfo{author}{\bibfnamefont{Y.}~\bibnamefont{Iseri}},
  \bibinfo{author}{\bibfnamefont{H.}~\bibnamefont{Kameyama}},
  \bibinfo{author}{\bibfnamefont{M.}~\bibnamefont{Kamimura}}, \bibnamefont{and}
  \bibinfo{author}{\bibfnamefont{M.}~\bibnamefont{Kawai}},
  \bibinfo{journal}{Prog.\ Theor.\ Phys.\ Suppl.\ {\bf 89}, 32}
  (\bibinfo{year}{1986}).

\bibitem[{\citenamefont{Austern et~al.}(1987)\citenamefont{Austern, Iseri,
  Kamimura, Kawai, Rawitscher, and Yahiro}}]{Aus87}
\bibinfo{author}{\bibfnamefont{N.}~\bibnamefont{Austern}},
  \bibinfo{author}{\bibfnamefont{Y.}~\bibnamefont{Iseri}},
  \bibinfo{author}{\bibfnamefont{M.}~\bibnamefont{Kamimura}},
  \bibinfo{author}{\bibfnamefont{M.}~\bibnamefont{Kawai}},
  \bibinfo{author}{\bibfnamefont{G.}~\bibnamefont{Rawitscher}},
  \bibnamefont{and} \bibinfo{author}{\bibfnamefont{M.}~\bibnamefont{Yahiro}},
  \bibinfo{journal}{Phys. Rep.} \textbf{\bibinfo{volume}{154}},
  \bibinfo{pages}{125} (\bibinfo{year}{1987}).

\bibitem[{\citenamefont{{Matsumoto {\it et al.}}}(2003)}]{Mat03}
\bibinfo{author}{\bibfnamefont{T.}~\bibnamefont{{Matsumoto {\it et al.}}}},
  \bibinfo{journal}{Phys. Rev.} \textbf{\bibinfo{volume}{C 68}},
  \bibinfo{pages}{064607} (\bibinfo{year}{2003}).

\bibitem[{\citenamefont{Mac\'{\i}as et~al.}(1987)\citenamefont{Mac\'{\i}as,
  Mart\'{\i}n, and {Y\'a\~nez}}}]{Mac87a}
\bibinfo{author}{\bibfnamefont{A.}~\bibnamefont{Mac\'{\i}as}},
  \bibinfo{author}{\bibfnamefont{F.}~\bibnamefont{Mart\'{\i}n}},
  \bibnamefont{and}
  \bibinfo{author}{\bibfnamefont{M.}~\bibnamefont{{Y\'a\~nez}}},
  \bibinfo{journal}{Phys. Rev.} \textbf{\bibinfo{volume}{A 36}},
  \bibinfo{pages}{4179} (\bibinfo{year}{1987}).

\bibitem[{\citenamefont{Bray}(1995)}]{Bra95}
\bibinfo{author}{\bibfnamefont{I.}~\bibnamefont{Bray}}, \bibinfo{journal}{Comp.
  Phys. Comm.} \textbf{\bibinfo{volume}{85}}, \bibinfo{pages}{1}
  (\bibinfo{year}{1995}).

\bibitem[{\citenamefont{Kuruoglu and Levin}(1982)}]{Kur82}
\bibinfo{author}{\bibfnamefont{Z.~C.} \bibnamefont{Kuruoglu}} \bibnamefont{and}
  \bibinfo{author}{\bibfnamefont{F.~S.} \bibnamefont{Levin}},
  \bibinfo{journal}{Phys. Rev. Lett.} \textbf{\bibinfo{volume}{48}},
  \bibinfo{pages}{899} (\bibinfo{year}{1982}).

\bibitem[{\citenamefont{Kuruoglu}(1991)}]{Kur91}
\bibinfo{author}{\bibfnamefont{Z.}~\bibnamefont{Kuruoglu}},
  \bibinfo{journal}{Phys. Rev.} \textbf{\bibinfo{volume}{C 43}},
  \bibinfo{pages}{1061} (\bibinfo{year}{1991}).

\bibitem[{\citenamefont{Hiyama et~al.}(2003)\citenamefont{Hiyama, Kino, and
  Kamimura}}]{Hiy03}
\bibinfo{author}{\bibfnamefont{E.}~\bibnamefont{Hiyama}},
  \bibinfo{author}{\bibfnamefont{Y.}~\bibnamefont{Kino}}, \bibnamefont{and}
  \bibinfo{author}{\bibfnamefont{M.}~\bibnamefont{Kamimura}},
  \bibinfo{journal}{Prog. Part. Nucl. Phys.} \textbf{\bibinfo{volume}{51}},
  \bibinfo{pages}{223} (\bibinfo{year}{2003}).

\bibitem[{\citenamefont{Matsumoto et~al.}(2004)}]{Mat04}
\bibinfo{author}{\bibfnamefont{T.}~\bibnamefont{Matsumoto}}
  \bibnamefont{et~al.}, \bibinfo{journal}{Phys. Rev.}
  \textbf{\bibinfo{volume}{C70}}, \bibinfo{pages}{061601}
  (\bibinfo{year}{2004}).

\bibitem[{\citenamefont{Petkov and Stoitsov}(1981)}]{Pet81}
\bibinfo{author}{\bibfnamefont{I.}~\bibnamefont{Petkov}} \bibnamefont{and}
  \bibinfo{author}{\bibfnamefont{M.}~\bibnamefont{Stoitsov}},
  \bibinfo{journal}{C.R. Acad. Bulg. Sci.} \textbf{\bibinfo{volume}{34}},
  \bibinfo{pages}{1651} (\bibinfo{year}{1981}).

\bibitem[{\citenamefont{P\'erez-Bernal
  et~al.}(2001{\natexlab{a}})\citenamefont{P\'erez-Bernal, Martel, Arias, and
  Gómez-Camacho}}]{Per01a}
\bibinfo{author}{\bibfnamefont{F.}~\bibnamefont{P\'erez-Bernal}},
  \bibinfo{author}{\bibfnamefont{I.}~\bibnamefont{Martel}},
  \bibinfo{author}{\bibfnamefont{J.~M.} \bibnamefont{Arias}}, \bibnamefont{and}
  \bibinfo{author}{\bibfnamefont{J.}~\bibnamefont{Gómez-Camacho}},
  \bibinfo{journal}{Phys. Rev. {\bf A63}, 052111}
  (\bibinfo{year}{2001}{\natexlab{a}}).

\bibitem[{\citenamefont{P\'erez-Bernal
  et~al.}(2001{\natexlab{b}})\citenamefont{P\'erez-Bernal, Martel, Arias, and
  G\'omez-Camacho}}]{Per01b}
\bibinfo{author}{\bibfnamefont{F.}~\bibnamefont{P\'erez-Bernal}},
  \bibinfo{author}{\bibfnamefont{I.}~\bibnamefont{Martel}},
  \bibinfo{author}{\bibfnamefont{J.~M.} \bibnamefont{Arias}}, \bibnamefont{and}
  \bibinfo{author}{\bibfnamefont{J.}~\bibnamefont{G\'omez-Camacho}},
  \bibinfo{journal}{Few Body Sys. Supl. {\bf 13}, 213}
  (\bibinfo{year}{2001}{\natexlab{b}}).

\bibitem[{\citenamefont{Rodr\'iguez-Gallardo
  et~al.}(2004)\citenamefont{Rodr\'iguez-Gallardo, Arias, and
  G\'omez-Camacho}}]{Rod04}
\bibinfo{author}{\bibfnamefont{M.}~\bibnamefont{Rodr\'iguez-Gallardo}},
  \bibinfo{author}{\bibfnamefont{J.~M.} \bibnamefont{Arias}}, \bibnamefont{and}
  \bibinfo{author}{\bibfnamefont{J.}~\bibnamefont{G\'omez-Camacho}},
  \bibinfo{journal}{Phys. Rev.} \textbf{\bibinfo{volume}{C69}},
  \bibinfo{pages}{034308} (\bibinfo{year}{2004}).

\bibitem[{\citenamefont{Moro et~al.}(2002)\citenamefont{Moro, Arias,
  G\'omez-Camacho, Martel, P\'erez-Bernal, Nunes, and Crespo}}]{Mor02}
\bibinfo{author}{\bibfnamefont{A.~M.} \bibnamefont{Moro}},
  \bibinfo{author}{\bibfnamefont{J.~M.} \bibnamefont{Arias}},
  \bibinfo{author}{\bibfnamefont{J.}~\bibnamefont{G\'omez-Camacho}},
  \bibinfo{author}{\bibfnamefont{I.}~\bibnamefont{Martel}},
  \bibinfo{author}{\bibfnamefont{F.}~\bibnamefont{P\'erez-Bernal}},
  \bibinfo{author}{\bibfnamefont{F.}~\bibnamefont{Nunes}}, \bibnamefont{and}
  \bibinfo{author}{\bibfnamefont{R.}~\bibnamefont{Crespo}},
  \bibinfo{journal}{Phys. Rev. {\bf C65}, 011602}  (\bibinfo{year}{2002}).

\bibitem[{\citenamefont{Martel et~al.}(2002)\citenamefont{Martel,
  P\'erez-Bernal, Rodr\'iguez-Gallardo, Arias, and Gómez-Camacho}}]{Mar02}
\bibinfo{author}{\bibfnamefont{I.}~\bibnamefont{Martel}},
  \bibinfo{author}{\bibfnamefont{F.}~\bibnamefont{P\'erez-Bernal}},
  \bibinfo{author}{\bibfnamefont{M.}~\bibnamefont{Rodr\'iguez-Gallardo}},
  \bibinfo{author}{\bibfnamefont{J.~M.} \bibnamefont{Arias}}, \bibnamefont{and}
  \bibinfo{author}{\bibfnamefont{J.}~\bibnamefont{Gómez-Camacho}},
  \bibinfo{journal}{Phys. Rev.} \textbf{\bibinfo{volume}{A65}},
  \bibinfo{pages}{052708} (\bibinfo{year}{2002}).

\bibitem[{\citenamefont{Aksough~{\it et al.}}(2003)}]{Aks03}
\bibinfo{author}{\bibfnamefont{F.}~\bibnamefont{Aksough~{\it et al.}}},
  \bibinfo{journal}{Review of the University of Milano, Ricerca Scientifica ed
  educazione permanente} \textbf{\bibinfo{volume}{Suppl. 122}}
  (\bibinfo{year}{2003}).

\bibitem[{\citenamefont{Egelhof}(2002)}]{Ege02}
\bibinfo{author}{\bibfnamefont{P.}~\bibnamefont{Egelhof}},
  \bibinfo{journal}{Nucl. Phys.} \textbf{\bibinfo{volume}{A 722}},
  \bibinfo{pages}{C254} (\bibinfo{year}{2002}).

\bibitem[{\citenamefont{Aumann et~al.}(1998)\citenamefont{Aumann, Chulkov,
  Pribora, and Smedberg}}]{Aum98}
\bibinfo{author}{\bibfnamefont{T.}~\bibnamefont{Aumann}},
  \bibinfo{author}{\bibfnamefont{L.~V.} \bibnamefont{Chulkov}},
  \bibinfo{author}{\bibfnamefont{V.~N.} \bibnamefont{Pribora}},
  \bibnamefont{and} \bibinfo{author}{\bibfnamefont{M.~H.}
  \bibnamefont{Smedberg}}, \bibinfo{journal}{Nucl. Phys.}
  \textbf{\bibinfo{volume}{A640}}, \bibinfo{pages}{24} (\bibinfo{year}{1998}).

\bibitem[{\citenamefont{{Aumann {\it et al.}}}(1999)}]{Aum99}
\bibinfo{author}{\bibfnamefont{T.}~\bibnamefont{{Aumann {\it et al.}}}},
  \bibinfo{journal}{Phys. Rev.} \textbf{\bibinfo{volume}{C 59}},
  \bibinfo{pages}{1252} (\bibinfo{year}{1999}).

\bibitem[{\citenamefont{{Aguilera {\em et al.}}}(2001)}]{Agu01}
\bibinfo{author}{\bibfnamefont{E.~F.} \bibnamefont{{Aguilera {\em et al.}}}},
  \bibinfo{journal}{Phys. Rev. C} \textbf{\bibinfo{volume}{63}},
  \bibinfo{pages}{061603} (\bibinfo{year}{2001}).

\bibitem[{\citenamefont{{Aguilera {\em et al.}}}(2000)}]{Agu00}
\bibinfo{author}{\bibfnamefont{E.~F.} \bibnamefont{{Aguilera {\em et al.}}}},
  \bibinfo{journal}{Phys. Rev. Lett.} \textbf{\bibinfo{volume}{84}},
  \bibinfo{pages}{5058} (\bibinfo{year}{2000}).

\bibitem[{\citenamefont{{Kakuee {\it et al.}}}(2003)}]{Kak03}
\bibinfo{author}{\bibfnamefont{O.~R.} \bibnamefont{{Kakuee {\it et al.}}}},
  \bibinfo{journal}{Nucl. Phys.} \textbf{\bibinfo{volume}{A 728}},
  \bibinfo{pages}{339} (\bibinfo{year}{2003}).

\bibitem[{\citenamefont{Zhukov et~al.}(1993)\citenamefont{Zhukov, Danilin,
  Fedorov, Bang, Thompson, and Vaagen}}]{Zhu93}
\bibinfo{author}{\bibfnamefont{M.~V.} \bibnamefont{Zhukov}},
  \bibinfo{author}{\bibfnamefont{B.~V.} \bibnamefont{Danilin}},
  \bibinfo{author}{\bibfnamefont{D.~V.} \bibnamefont{Fedorov}},
  \bibinfo{author}{\bibfnamefont{J.~M.} \bibnamefont{Bang}},
  \bibinfo{author}{\bibfnamefont{I.~J.} \bibnamefont{Thompson}},
  \bibnamefont{and} \bibinfo{author}{\bibfnamefont{J.~S.}
  \bibnamefont{Vaagen}}, \bibinfo{journal}{Physics Reports}
  \textbf{\bibinfo{volume}{231}}, \bibinfo{pages}{151} (\bibinfo{year}{1993}).

\bibitem[{\citenamefont{Hiyama and Kamimura}(1995)}]{Hi95}
\bibinfo{author}{\bibfnamefont{E.}~\bibnamefont{Hiyama}} \bibnamefont{and}
  \bibinfo{author}{\bibfnamefont{M.}~\bibnamefont{Kamimura}},
  \bibinfo{journal}{Nucl. Phys.} \textbf{\bibinfo{volume}{A588}},
  \bibinfo{pages}{35c} (\bibinfo{year}{1995}).

\bibitem[{\citenamefont{Danilin et~al.}(1997)\citenamefont{Danilin, Rogde,
  Ershov, Heiberg-Andersen, Vaagen, Thompson, and Zhukov}}]{Dan97}
\bibinfo{author}{\bibfnamefont{B.~V.} \bibnamefont{Danilin}},
  \bibinfo{author}{\bibfnamefont{T.}~\bibnamefont{Rogde}},
  \bibinfo{author}{\bibfnamefont{S.~N.} \bibnamefont{Ershov}},
  \bibinfo{author}{\bibfnamefont{H.}~\bibnamefont{Heiberg-Andersen}},
  \bibinfo{author}{\bibfnamefont{J.~S.} \bibnamefont{Vaagen}},
  \bibinfo{author}{\bibfnamefont{I.~J.} \bibnamefont{Thompson}},
  \bibnamefont{and} \bibinfo{author}{\bibfnamefont{M.~V.}
  \bibnamefont{Zhukov}}, \bibinfo{journal}{Phys. Rev.}
  \textbf{\bibinfo{volume}{C 55}}, \bibinfo{pages}{R577}
  (\bibinfo{year}{1997}).

\bibitem[{\citenamefont{Fedorov et~al.}(2003)\citenamefont{Fedorov, Garrido,
  and Jensen}}]{Fed03}
\bibinfo{author}{\bibfnamefont{D.~V.} \bibnamefont{Fedorov}},
  \bibinfo{author}{\bibfnamefont{E.}~\bibnamefont{Garrido}}, \bibnamefont{and}
  \bibinfo{author}{\bibfnamefont{A.~S.} \bibnamefont{Jensen}},
  \bibinfo{journal}{Few Body Syst.} \textbf{\bibinfo{volume}{33}},
  \bibinfo{pages}{153} (\bibinfo{year}{2003}).

\bibitem[{\citenamefont{Garrido et~al.}(2005)\citenamefont{Garrido, Fedorov,
  Jensen, and Fynbo}}]{Gar04}
\bibinfo{author}{\bibfnamefont{E.}~\bibnamefont{Garrido}},
  \bibinfo{author}{\bibfnamefont{D.~V.} \bibnamefont{Fedorov}},
  \bibinfo{author}{\bibfnamefont{A.~S.} \bibnamefont{Jensen}},
  \bibnamefont{and} \bibinfo{author}{\bibfnamefont{H.~O.~U.}
  \bibnamefont{Fynbo}}, \bibinfo{journal}{Nucl. Phys.}
  \textbf{\bibinfo{volume}{A748}}, \bibinfo{pages}{39} (\bibinfo{year}{2005}).

\bibitem[{\citenamefont{Myo et~al.}(2001)\citenamefont{Myo, Kato, Aoyama, and
  Ikeda}}]{Myo01}
\bibinfo{author}{\bibfnamefont{T.}~\bibnamefont{Myo}},
  \bibinfo{author}{\bibfnamefont{K.}~\bibnamefont{Kato}},
  \bibinfo{author}{\bibfnamefont{S.}~\bibnamefont{Aoyama}}, \bibnamefont{and}
  \bibinfo{author}{\bibfnamefont{K.}~\bibnamefont{Ikeda}},
  \bibinfo{journal}{Phys. Rev.} \textbf{\bibinfo{volume}{C63}},
  \bibinfo{pages}{054313} (\bibinfo{year}{2001}).

\bibitem[{\citenamefont{Suzuki}(1991)}]{Su91}
\bibinfo{author}{\bibfnamefont{Y.}~\bibnamefont{Suzuki}},
  \bibinfo{journal}{Nucl. Phys.} \textbf{\bibinfo{volume}{A528}},
  \bibinfo{pages}{395} (\bibinfo{year}{1991}).

\bibitem[{\citenamefont{Funada et~al.}(1994)\citenamefont{Funada, Kameyama, and
  Sakuragi}}]{Fu94}
\bibinfo{author}{\bibfnamefont{S.}~\bibnamefont{Funada}},
  \bibinfo{author}{\bibfnamefont{H.}~\bibnamefont{Kameyama}}, \bibnamefont{and}
  \bibinfo{author}{\bibfnamefont{Y.}~\bibnamefont{Sakuragi}},
  \bibinfo{journal}{Nucl. Phys.} \textbf{\bibinfo{volume}{A575}},
  \bibinfo{pages}{93} (\bibinfo{year}{1994}).

\bibitem[{\citenamefont{Navratil and Barrett}(1996)}]{Na96}
\bibinfo{author}{\bibfnamefont{P.}~\bibnamefont{Navratil}} \bibnamefont{and}
  \bibinfo{author}{\bibfnamefont{B.~R.} \bibnamefont{Barrett}},
  \bibinfo{journal}{Phys. Rev.} \textbf{\bibinfo{volume}{C 54}},
  \bibinfo{pages}{2986} (\bibinfo{year}{1996}).

\bibitem[{\citenamefont{Cs\'ot\'o}(1993)}]{Cso93}
\bibinfo{author}{\bibfnamefont{A.}~\bibnamefont{Cs\'ot\'o}},
  \bibinfo{journal}{Phys. Rev.} \textbf{\bibinfo{volume}{C 48}},
  \bibinfo{pages}{165} (\bibinfo{year}{1993}).

\bibitem[{\citenamefont{Wurzer and Hofmann}(1997)}]{Wur97}
\bibinfo{author}{\bibfnamefont{J.}~\bibnamefont{Wurzer}} \bibnamefont{and}
  \bibinfo{author}{\bibfnamefont{H.~M.} \bibnamefont{Hofmann}},
  \bibinfo{journal}{Phys. Rev.} \textbf{\bibinfo{volume}{C 55}},
  \bibinfo{pages}{688} (\bibinfo{year}{1997}).

\bibitem[{\citenamefont{Thompson et~al.}(2004)\citenamefont{Thompson, Nunes,
  and Danilin}}]{face}
\bibinfo{author}{\bibfnamefont{I.~J.} \bibnamefont{Thompson}},
  \bibinfo{author}{\bibfnamefont{F.~M.} \bibnamefont{Nunes}}, \bibnamefont{and}
  \bibinfo{author}{\bibfnamefont{B.~V.} \bibnamefont{Danilin}},
  \bibinfo{journal}{Comput. Phys. Commun.} \textbf{\bibinfo{volume}{161}},
  \bibinfo{pages}{87} (\bibinfo{year}{2004}).

\bibitem[{\citenamefont{Bang and Gignoux}(1979)}]{Bang79}
\bibinfo{author}{\bibfnamefont{J.}~\bibnamefont{Bang}} \bibnamefont{and}
  \bibinfo{author}{\bibfnamefont{C.}~\bibnamefont{Gignoux}},
  \bibinfo{journal}{Nucl. Phys.} \textbf{\bibinfo{volume}{A 313}},
  \bibinfo{pages}{119} (\bibinfo{year}{1979}).

\bibitem[{\citenamefont{Thompson et~al.}(2000)\citenamefont{Thompson, Danilin,
  Efros, Vaagen, Bang, and Zhukov}}]{Tho00}
\bibinfo{author}{\bibfnamefont{I.~J.} \bibnamefont{Thompson}},
  \bibinfo{author}{\bibfnamefont{B.~V.} \bibnamefont{Danilin}},
  \bibinfo{author}{\bibfnamefont{V.~D.} \bibnamefont{Efros}},
  \bibinfo{author}{\bibfnamefont{J.~S.} \bibnamefont{Vaagen}},
  \bibinfo{author}{\bibfnamefont{J.~M.} \bibnamefont{Bang}}, \bibnamefont{and}
  \bibinfo{author}{\bibfnamefont{M.~V.} \bibnamefont{Zhukov}},
  \bibinfo{journal}{Phys. Rev.} \textbf{\bibinfo{volume}{C 61}},
  \bibinfo{pages}{24318} (\bibinfo{year}{2000}).

\bibitem[{\citenamefont{Gogny et~al.}(1970)\citenamefont{Gogny, Pires, and
  de~Tourreil}}]{gpt}
\bibinfo{author}{\bibfnamefont{D.}~\bibnamefont{Gogny}},
  \bibinfo{author}{\bibfnamefont{P.}~\bibnamefont{Pires}}, \bibnamefont{and}
  \bibinfo{author}{\bibfnamefont{R.}~\bibnamefont{de~Tourreil}},
  \bibinfo{journal}{Phys. Lett.} \textbf{\bibinfo{volume}{32B}},
  \bibinfo{pages}{591} (\bibinfo{year}{1970}).

\bibitem[{\citenamefont{Crespo et~al.}(2002)\citenamefont{Crespo, Thompson, and
  Korsheninnikov}}]{Cre02}
\bibinfo{author}{\bibfnamefont{R.}~\bibnamefont{Crespo}},
  \bibinfo{author}{\bibfnamefont{I.~J.} \bibnamefont{Thompson}},
  \bibnamefont{and} \bibinfo{author}{\bibfnamefont{A.~A.}
  \bibnamefont{Korsheninnikov}}, \bibinfo{journal}{Phys. Rev.}
  \textbf{\bibinfo{volume}{C66}}, \bibinfo{pages}{021002}
  (\bibinfo{year}{2002}).

\bibitem[{\citenamefont{Brink and Satchler}(1994)}]{BS94}
\bibinfo{author}{\bibfnamefont{D.~M.} \bibnamefont{Brink}} \bibnamefont{and}
  \bibinfo{author}{\bibfnamefont{G.~R.} \bibnamefont{Satchler}},
  \emph{\bibinfo{title}{Angular Momentum}} (\bibinfo{publisher}{Clarendon,
  Oxford}, \bibinfo{year}{1994}).

\end{thebibliography}

\end{document}